\documentclass[runningheads]{llncs}

\usepackage[pdftex]{graphicx}
\usepackage{hyperref}
\usepackage{cite}

\usepackage{amsmath}
\usepackage{amsfonts}
\usepackage[linesnumbered,ruled,noend]{algorithm2e}
\usepackage{caption}
\usepackage{subcaption}
\usepackage{booktabs}
\usepackage{colortbl}
\usepackage{listings}
\usepackage{url}
\usepackage{tikz}
 
\usepackage{pifont}%

\newcommand{\parhead}[1]{\textbf{#1.}}

\usepackage{todonotes}
\usepackage{marginnote}
\newcounter{researchquestion}

\newcounter{challenge}

\makeatletter
\newcommand{\researchquestionautorefname}{RQ\@gobble}
\newcommand{\challengeautorefname}{C\@gobble}
\makeatother

\usepackage{paralist} %
\usepackage{multirow}
\usepackage{xurl}

\usepackage[nolist,nohyperlinks]{acronym}
\makeatletter
\newcommand*{\iacl}[1]{\@iaci{#1} \acl{#1}}
\newcommand*{\iaclp}[1]{\@iaci{#1} \aclp{#1}}
\newcommand*{\iacs}[1]{\@iaci{#1} \acs{#1}}
\newcommand*{\iacsp}[1]{\@iaci{#1} \acsp{#1}}
\newcommand*{\iacf}[1]{\@iaci{#1} \acf{#1}}
\newcommand*{\iacfp}[1]{\@iaci{#1} \acfp{#1}}

\newcommand*{\Iacl}[1]{\@firstupper{\iacl{#1}}}
\newcommand*{\Iaclp}[1]{\@firstupper{\iaclp{#1}}}
\newcommand*{\Iacs}[1]{\@firstupper{\iacs{#1}}}
\newcommand*{\Iacsp}[1]{\@firstupper{\iacsp{#1}}}
\newcommand*{\Iacf}[1]{\@firstupper{\iacf{#1}}}
\newcommand*{\Iacfp}[1]{\@firstupper{\iacfp{#1}}}
\makeatother

\begin{acronym}
	\acro{graph}[EDG]{Execution Divergence Graph}
	\acroindefinite{graph}{an}{an}
	\acro{cfg}[CFG]{Control Flow Graph}
\end{acronym}

\begin{document}

 \title{Execution Divergence Graphs: Effective Novelty Detection from Execution Traces}
 \title{Execution Divergence Graphs: Runtime-Instrumentation Feedback Mechanism}
 \title{Execution Divergence Graphs:\\ Effective Discovery of Control-Flows from Execution Traces as Fuzzing Feedback}

\titlerunning{Execution Divergence Graphs}

\author{
Yu-De Lin\and
Nils Ole Tippenhauer
} 

\institute{
CISPA Helmholtz Center for Information Security, Germany}

\maketitle

\begin{abstract}
Fuzz testing is a popular approach to the security testing of proprietary software. 
Efficient testing strategies rely on execution feedback to guide the input generation process, particularly when the basic blocks in the binary can be directly observed and instrumented. Unfortunately, collecting such feedback is impossible in scenarios such as in-situ fuzzing of black-box devices and the fuzzing of obfuscated compiled binaries. In this work, we discuss approaches to guide the fuzzer using feedback derived from a control-flow-graph-like (CFG-like) structure constructed from runtime execution.

We start by outlining a simple divergence-detection approach that identifies unique execution traces, and then present an improved approach based on an Execution Divergence Graph (EDG).
We implement both approaches and demonstrate that they outperform a baseline blind fuzzer. 
In addition, we discuss particular challenges, such as repeated code execution in loops, and show that the EDG-based approach handles them effectively.
We then demonstrate that our approach enables effective fuzzing of a number of obfuscated targets, and compare its performance in scenarios where static instrumentation is impossible. 
While we focus on a scenario in which full instruction traces are directly observable by the attacker, our scheme can also be applied in scenarios with other feedback channels, such as power consumption.

\end{abstract}

\section{Introduction}\label{sec:intro}

Fuzzing is an automated software-testing technology that works by injecting random inputs (called ``fuzz''~\cite{Miller1990}) into a program to expose bugs, crashes, or security vulnerabilities. Unguided fuzzers generate inputs from a corpus~\cite{zzuf,radamsa} or predefined-input rules~\cite{peach} to explore simple programs. However, unguided fuzzers struggle to reach deeper code paths in complex programs, where serious vulnerabilities are often located. To address this issue, modern guided test approaches~\cite{afl-whitepaper,libafl,tfuzz,angora,SchumiloAGSH17,AschermannSBGH19} use coverage, which represents how many code paths are visited during testing, to guide the fuzzer. This coverage-guided fuzzing requires static instrumentation to generate feedback.

Instrumentation, however, has several limitations. First, it requires access to the target's source code or binary for successful code modification. Even when a binary is available, its author can use tools to obfuscate it, rendering instrumentation ineffective. For instance, obfuscators such as Movfuscator~\cite{movfuscator2025} and Tigress~\cite{tigress} can hide all standard branch instructions. Without distinct control-flow instructions like branches, the concept of a basic block is lost, and the fuzzer cannot accurately map or track connections between basic blocks (edges). Consequently, the fuzzer cannot obtain the crucial control-flow information, such as which paths an input has visited, needed to guide its exploration.

In this work, we present several approaches to address this issue, and show how to generate suitable feedback for fuzzing by analyzing target execution traces. 
Such traces could, for example, consist of a sequence of the executed instruction addresses without further annotation. 
Even without static analysis of the target binary, we show how to process such execution traces to detect execution of novel code segments, while maintaining computational and memory efficiency.
While naive approaches to this problem simply store prior execution traces and compare them to new traces, we show that such approaches are particularly unable to detect repeated execution of code segments, and thus vastly overestimate the novelty of inputs. 
To address this, we introduce the concept of Execution Divergence Graphs (EDGs), a CFG-like structure for tracking observed execution sequences and describing the target program. Using an EDG to track input executions allows valuable coverage feedback to be generated for standard fuzzers. 
In our experiments, we demonstrate that our approach achieves performance similar to classic instrumentation-guided fuzzing, without requiring explicit instrumentation of the target binary.

Our contributions are as follows:

\begin{itemize}
    \item We demonstrate that the divergence between execution traces can guide a coverage-guided fuzzer.
    \item We introduce a graph structure based on execution traces to improve the execution-divergence feedback given to the fuzzer.
    \item We implement our approach and evaluate it on several real-world programs to demonstrate that our approach is applicable to standard fuzzing tasks.
    \item We also demonstrate our approach on several obfuscated targets (where classic instrumentation is impossible), and show that our approach can still guide the fuzzer to explore the obfuscated programs efficiently.  
    \item Our artifact can be found at \url{https://anonymous.4open.science/r/EDG-BBC4}.
\end{itemize}

\section{Background}

\subsection{Coverage-Guided Fuzzing}
The ultimate objective of a coverage-guided fuzzer is to generate inputs such that, by the end of a fuzzing campaign, every possible control-flow path has been traversed at least once.
Coverage-guided fuzzers usually use instrumentation to extract the control flow of the target for a given input. Based on this feedback, they attempt to generate additional inputs that exercise different paths from those taken by previous inputs, thus maximizing code coverage.

Coverage is often measured in the form of \emph{edge coverage}~\cite{afl-whitepaper}. The target binary is instrumented so that it emits an edge-coverage hash map for each input. This map expresses the transitions between basic blocks that occur at runtime in a condensed form. The fuzzer maintains a cumulative hash map to track all known edges and their frequencies of occurrence. Comparing the hash map of a single execution run (using one particular input) with the cumulative hash map allows the fuzzer to assess whether new edges were discovered or whether known edges were traversed more or less frequently.

\subsection{Guided Black-box Fuzzing}

Although coverage-guided fuzzing effectively finds bugs in programs that support post-instrumentation edge feedback, it is inapplicable in various scenarios, such as network and protocol fuzzing, where static instrumentation is not feasible. Therefore, black-box fuzzing that does not leverage the internal information of a program is widely applied in these scenarios.

However, conventional black-box fuzzing approaches, such as mutation-based and generation-based fuzzing, are not sufficient in these scenarios. Thus, these approaches design their own feedback mechanism based on the externally observable states such as responses from IoT devices, protocol states, and messages between network nodes~\cite{kim2023intender,zou2025distfuzz,deruiter2015tls} to \emph{guide} black-box fuzzing.

\section{Novelty Detection from Execution Traces}

\subsection{System Model}

\begin{figure}[tb]
    \centering
    \includegraphics[width=0.5\linewidth]{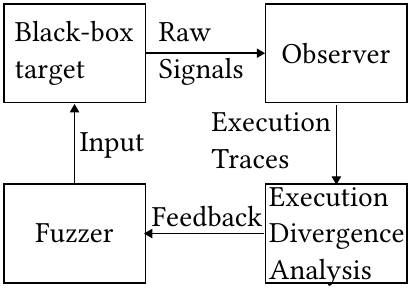}
    \caption{System Overview}
    \label{fig:sys_model}
\end{figure}

We describe the components illustrated in Figure~\ref{fig:sys_model} of the system model. 

\textbf{Black-box Target} is a system or device that accepts input for execution. A major challenge arises when the binary is inaccessible. This prevents third-party testers from directly recovering the underlying code, which is a prerequisite for effective software testing, especially coverage-guided fuzzing.

\textbf{Observer} is an external device (e.g., J-Trace Debugger or an oscilloscope) or software (e.g., GDB) that captures the \textit{raw signals} from a black-box target. The observer decodes the raw signals into the \textit{execution traces}, which record the time-series information flow during program execution. The data points vary by observer type: if the observer is a debugging tool, the trace records text indicating which instruction or function was executed at a specific timestamp. Alternatively, if the observer is a measurement device, such as an oscilloscope monitoring power signals, the time-series data consists of floating-point values representing power consumption.

\textbf{Execution Divergence Analysis} Component is designed to identify differences between execution traces from a pair of program executions. It executes the \textit{Divergence Detection} process to quantify the difference between two execution traces ($T_1$ and $T_2$), using an element-wise heuristic function $H(T_1, T_2) = v$, which yields a difference vector $v$ aligned in time with the traces. A divergence point is then designated at any timestamp $s$ where the corresponding value $v_s$ exceeds a predefined threshold $\tau$ ($v_s > \tau$), indicating the time at which the observed divergence between the traces begins. The component then leverages the divergence results to create \textit{trace segments}. If the trace segments are sequential instructions, we refer to these trace segments as \textit{code segments}. \textbf{Fuzzer} uses the trace segments as feedback from the black-box target and generates inputs accordingly. 

\parhead{The Fuzzing Loop} The pseudocode of the fuzzing loop is shown in Algorithm~\ref{alg:overall_fuzz_loop}. The fuzzer first generates input $i$ from its given corpus without feedback (line 2). The generated input $i$ is then sent to the black-box target connected to the observer for processing (line 4). When the processing is complete, the observer obtains the trace information $t$ for the execution divergence analysis module (line 5). Subsequently, the module returns the analysis result as feedback to the fuzzer, which then generates the next input accordingly (line 6). If the given edge feedback is deemed interesting by the fuzzer, the input $i$ is added to the corpus $\mathbb{C}$.

\begin{algorithm}
	\SetAlgoLined
	\KwIn{$\mathbb{C}$: the seeded corpus for the fuzzer}
	\KwData{$\mathbb{T}$: previously observed unique traces; $i$: input; $f$: feedback provided by $DivergenceAnalysis(.)$; $t$: the observed execution trace.}
    
    $\mathbb{T} \leftarrow \emptyset$\;
    $i,\mathbb{C} \leftarrow Fuzzer(\emptyset, \mathbb{C})$\tcp*[r]{Generate an initial input without feedback.}
    \While{\texttt{true}}{
        $t \leftarrow Observer(Blackbox(i))$\;
        $f,\mathbb{T} \leftarrow DivergenceAnalysis(t, \mathbb{T})$\;
        $i,\mathbb{C} \leftarrow Fuzzer(f, \mathbb{C})$\;
    }
	\caption{Overall fuzzing loop with execution divergence feedback.}
	\label{alg:overall_fuzz_loop}
\end{algorithm}

\subsection{Research Questions}

\begin{itemize}
\item \parhead{RQ1}\label{rq-edg-efficient} Given a new execution trace and all previous traces, how can we efficiently determine whether the new execution trace follows a previously known code path (and identify the matching old trace(s))?

\item \parhead{RQ2}\label{rq-novelty-discovery}
Given a new execution trace and all previous traces, how can we efficiently determine whether the new trace contains novel code segments or novel transitions?

\item \parhead{RQ3}\label{rq-feedback} How can novel code and transition discovery be used as feedback for the fuzzer?

\end{itemize}

\subsection{Simple Divergence Approach}\label{subsec:simplediv}

We start by addressing \textbf{RQ1}. The simplest approach for detecting novelty in a new execution trace relies on a divergence-detection function that compares the new trace against all previously recorded traces (line 2 in Alg~\ref{alg:simple-divergence}). 

We first calculate the vector $v$ using the heuristic function $H(.)$ (line 3) and check at which index the value $v_s$ exceeds the threshold $\tau$. This indicates that two traces start to diverge at $s$, and the comparison then continues with the next $T$ (line 6-7 and line 2). When the new trace $t$ has been compared with all traces in $\mathbb{T}$, we add the trace into the trace set $\mathbb{T}$ (line 9). Lastly, we return the $t$ and $\mathbb{T}$ to the Fuzzer (line 10).

In contrast, if there is no value higher than threshold, the new trace $t$ 
is a duplicate and we stop further comparison with the rest of the traces in $\mathbb{T}$. The matching trace $T$ and the trace set $\mathbb{T}$ are then returned to the fuzzer (line 8). This method deems a new execution trace to be novel only if it diverges from every single recorded trace.

Although simple to implement, this technique suffers from two performance-related drawbacks: \textbf{Recomparing Shared Traces} and the \textbf{False Novelty from Repeating Code Segments}.

\begin{algorithm}
	\SetAlgoLined
	\KwIn{$\mathbb{T}$: set of unique traces; $t$: the new trace; $\tau$ threshold for trace segment comparison}
    \KwResult{$\mathbb{T}$ and $t$}
	\KwData{$duplicate$: a boolean value to denote a new trace is identical to one of the traces in $\mathbb{T}$.}
    $duplicate \leftarrow $ \texttt{false}\;
    \For{$T$ in $\mathbb{T}$}{
        $v \leftarrow H(T,t)$\;
        \DontPrintSemicolon
        \For{$v_s$ in $v$ \tcp*[r]{Check if two traces differ}}{ 
        \PrintSemicolon
            \If{$v_s > \tau$ }{
                break\;
            }
        }
        \Return $t,\mathbb{T}$\;
    }
    \DontPrintSemicolon
        $\mathbb{T} \leftarrow \mathbb{T} \bigcup \{t\}$\;
	\Return $t, \mathbb{T}$
	\caption{Pseudocode of simple Divergence Analysis Approach.}
	\label{alg:simple-divergence}
\end{algorithm}

\textbf{The Recomparing Shared Traces} issue arises because the method redundantly checks common code segments shared among multiple recorded execution traces. Assuming that all $n$ recorded traces share the same code segment \texttt{ABCDE} in the beginning, the approach has to perform at least $n \times 5$ identical comparisons. This excessive and redundant computation drastically degrades performance, particularly as the number of the recorded traces grows.

\textbf{False Novelty from Repeating Code Segments} arises when code segments repeated through loop iterations are mistakenly classified as novel. For instance, a program contains a code segment \texttt{C} that checks whether a byte is legal or not. The program uses this code segment to validate input bytes, which may vary in number depending on input length. When an input length is 3 bytes, the code segment will repeat 3 times, creating execution trace \texttt{CCC}. If a new input is 5 bytes long, the new execution trace will become \texttt{CCCCC}. By comparing the two execution traces \texttt{CCC} and \texttt{CCCCC}, the approach concludes that the trace \texttt{CCCCC} is novel. However, this code segment has already been discovered.

\subsection{Execution Divergence Graph-based Approach} \label{subsec:edg}

To address \textbf{RQ2} and overcome the significant performance impact caused by the redundant recomparison of shared traces, we now introduce the Execution Divergence Graph (\emph{EDG})-based approach.
The EDG data structure enables merging identical code segments shared across different execution traces. Unlike the baseline approach, which stores each trace identified as divergent, the EDG approach merges common, identical code segments between new and existing traces before storage. For instance, given a new trace \texttt{ABCDEFG} and an existing trace \texttt{ABCDECCA} within the EDG, the divergence detection function first identifies the shared code segments \texttt{ABCDE}. The graph then factors out this shared segment and connects it to the remaining segments, \texttt{FG} and \texttt{CCA}. In this example, the newly created shared code segment \texttt{ABCDE} and the two remaining segments are novel. Eventually, even if the EDG contains $n$ different execution traces sharing the code segment \texttt{ABCDE}, any new incoming trace only needs to compare with that merged segment once, drastically improving efficiency.

\parhead{Repeated Code Segments}
However, the simple EDG cannot solve the false-novelty problems caused by repeating code segments. Consider a simple EDG initialized with a stored execution trace $T_{1}$=\texttt{ABAB}. When a new incoming execution trace $T_{new}$=\texttt{ABABAB} arrives, it triggers the graph to create a new trace $T_{2}$=\texttt{AB}, connected to $T_1$. The graph then marks $T_2$ as a novel code segment. Furthermore, when there is a new execution trace $T_{new}$=\texttt{ABABABAB}, the graph iteratively creates another code segment $T_3$=\texttt{AB}, connected to $T_2$ and again marks $T_3$ as novel. In fact, $T_2$ and $T_3$ are the repeating code segments \texttt{AB}, demonstrating the simple EDG's failure to recognize repeating code segments.

\parhead{Backlinking in EDG} Backlinking is a mechanism we introduced to resolve the false positives from repeating code segments in the graph. This process is initiated immediately when the divergence detection function identifies a divergence between the incoming trace and the stored trace, signaling a candidate new segment. Backlinking then broadcasts the candidate segments across the entire graph and searches the entire graph for any existing identical code segments.
Following the previous example, when the graph obtains $T_2$=\texttt{AB}, backlinking broadcasts $T_2$ and discovers that there is an exact match with its parent trace $T_1$=\texttt{ABAB}, leading to its merger with the identical part in $T_1$ and establishing a \textit{self-loop connection}. After refactoring the identical part, $T_1$ is reduced to \texttt{AB} and broadcast by backlinking. Another exact match then occurs between $T_1$ and $T_2$, causing $T_1$ to be merged once again with $T_2$. With backlinking, no matter how many times a new incoming trace repeats \texttt{AB}, the EDG always considers it a non-novel code segment.

\parhead{Execution Divergence Analysis with EDG} The new trace $t$ is first used to traverse EDG $\mathbb{T}$, which stores all previously observed traces (line 1 in Alg~\ref{alg:edg}). After traversal, EDG returns the edges $f$ traversed by the new trace and the unmatched residual part $t^\prime$. If the trace has a perfect match in the graph, then $t^\prime$ is empty, and the traversed edges $f$ are returned. Conversely, when $t^\prime$ is not empty, the EDG triggers $backlink(.)$ to integrate $t^\prime$ (line 3). Since the graph is updated after $backlink(.)$, and some nodes and edges change, we traverse the trace $t$ again to obtain the latest traversed edges (line 4). More details about $traverse(.)$ and $backlink(.)$ are provided in Section~\ref{subsec:match_first_traversal} and Section~\ref{subsec:backlink_candidates}, respectively.

\begin{algorithm}
	\SetAlgoLined
    \KwIn{$\mathbb{T}$: EDG, containing unique traces observed so far; $t$ is a new trace.}
    \KwResult{$f$: the traversed edges in EDG.}
	\KwData{$t^\prime$: the unmatched part of the new trace $t$ after the first $traverse(.)$.}
    $(t^\prime, f) \leftarrow traverse(\mathbb{T},t)$\;
    \If{$t^\prime$ is not $\emptyset$} {
        $\mathbb{T} \leftarrow backlink(\mathbb{T}, t^\prime)$\;
        $(\emptyset, f) \leftarrow traverse(\mathbb{T},t)$\;
    }
    \Return $f, \mathbb{T}$
	\caption{Pseudocode of \emph{EDG}-based Execution Divergence Analysis.}
	\label{alg:edg}
\end{algorithm}

\subsection{Generating Feedback from Execution Divergence Analysis}

\begin{table}[t]
    \centering
    \caption{An example demonstrates how \textit{simple-div} and \textit{EDG} convert execution traces into AFL feedback.}
    \label{tab:trace-segment-feedback}
    \begin{tabular}{l  c  c  c  c  c}
        \toprule
         \multirow{2}{*}{Trace} & \multicolumn{2}{c}{\textit{simple-div}} & \multicolumn{3}{c}{\textit{EDG}} \\
        \cmidrule(lr){2-3} \cmidrule(lr){4-6}
           & Feedback & Interesting & Segments & Feedback & Interesting \\
        \midrule
         \texttt{S} & \texttt{S}:1 & Yes & \texttt{S} & $\emptyset$:1 & Yes \\
         \texttt{SAB} & \texttt{SAB}:1 & Yes & \texttt{S;AB} & \texttt{S$\to$AB}:1 & Yes \\
         \texttt{SABABABAB} & \texttt{SABABABAB}:1 & Yes & \texttt{S;AB} & \texttt{S$\to$AB}:1,\texttt{AB$\to$AB}:3, & Yes \\
         \texttt{SABABABAB} & \texttt{SABABABAB}:1 & No & \texttt{S;AB} & \texttt{S$\to$AB}:1,\texttt{AB$\to$AB}:3, & No \\
         \texttt{SABABABABAB} & \texttt{SABABABABAB}:1 & Yes & \texttt{S;AB} & \texttt{S$\to$AB}:1,\texttt{AB$\to$AB}:4 & No \\ 
        \bottomrule
    \end{tabular}
\end{table}

\parhead{Overview} In this section, we address \textbf{RQ3} and discuss \emph{how, based on an EDG, we can guide an AFL fuzzer by execution traces while fuzzing an uninstrumented target}. We first explain how AFL's hit-count bucket mechanism determines whether feedback is interesting and how \textit{simple-div} and \textit{EDG} interact with AFL through feedback. We use five execution traces shown in Table~\ref{tab:trace-segment-feedback} as examples to describe how \textit{simple-div} and \textit{EDG} work together with AFL.

\parhead{AFL's Hit-Count Bucket} AFL does not preserve the exact execution count of each edge. Instead, it maps the count into a small set of hit-count buckets, such as 1, 2, 3--4, 5--8, 9--16, 17--32, 33--128, and 129 or more. This coarse-grained encoding reduces noise from minor count fluctuations while still capturing meaningful execution changes. Therefore, an input is considered interesting not only when it reaches a previously unseen edge, but also when it drives an already known edge into a bucket that has not been observed before. 

\parhead{Feedback from \textit{simple-div}} For each trace, \textit{simple-div} generates one edge feedback for AFL, always with a hit count of 1. Each individual trace is mapped to a different edge, while identical traces are mapped to the same edge. In our example, four edges are reported to AFL, corresponding to \texttt{S}, \texttt{SAB}, \texttt{SABABABAB}, and \texttt{SABABABABAB}. As the second occurrence of \texttt{SABABABAB} is identical to the first occurrence, it is reported as the same edge, again with a hit count of 1. Thus, AFL does not consider it interesting. In summary, AFL considers all distinct traces in this example interesting because they are reported as previously unseen edges, whereas the repeated occurrence of \texttt{SABABABAB} is ignored because it is reported in the same way as before. Based on the five example traces, we identify two explicit drawbacks of \textit{simple-div}: first, every divergent trace is treated similarly because all divergent traces are placed in bucket 1; second, the repeated segment \texttt{AB} causes repeated traces to be treated as novel simply because it appears more times.

\parhead{Feedback from \textit{EDG} Analysis} Unlike \emph{simple-div}, \emph{EDG} converts traces into a graph representation. In this graph, links between nodes (i.e., transitions between code blocks) are used as feedback. This is very similar to standard AFL feedback based on links in a CFG. The first trace is a special case for \emph{EDG} because the graph contains only a single node (\texttt{S}). The feedback reported to AFL is represented as a special edge 0 with a count of 1. When the second trace, \texttt{SAB}, arrives, our \emph{EDG} analysis adds the segment \texttt{AB} and the corresponding link between \texttt{S} and \texttt{AB} into the EDG. This link is then reported to AFL with a count of 1. The \emph{EDG} analysis does not find any new code segment in the third trace, \texttt{SABABABAB}, but it finds that the transition \texttt{AB$\to$AB} repeats three times in the trace. Since the transition \texttt{AB$\to$AB} is new and occurs three times in the execution, AFL places the new transition in bucket 3-4 and marks it as interesting. The fourth trace is identical to the third trace and is therefore not interesting. For the fifth trace, our \emph{EDG} analysis finds that the transition \texttt{AB$\to$AB} repeats four times, which is different from before. However, it remains in bucket 3-4 because the number of repetitions does not exceed the bucket cap of 4 in this case. Therefore, the fifth trace is not considered interesting by AFL. In summary, this example demonstrates that our \emph{EDG} analysis provides more informative feedback than the simple divergence detection approach.

\section{Detailed Design of Execution Divergence Graphs}\label{sec:edg_in_detail}

\subsection{Definition}

Unlike a control-flow graph, which captures all possible connections between basic blocks based on complete information available from an instrumented binary, the EDG is constructed solely from execution traces collected at runtime from a black-box target. It organizes these traces according to their observed divergences, enabling third-party testers to analyze the target’s behavior without access to its internal code structure. Because the EDG depends entirely on observed executions, it cannot reveal execution paths that never occur at runtime, even though such paths may be represented in a traditional control-flow graph. However, the EDG can capture concrete execution paths that static CFG recovery may miss, such as indirect function calls whose callees cannot be determined at compile time. Consequently, the EDG serves as a \textit{dynamic representation of a control-flow graph}, capturing the behavior that actually manifests during execution.

\subsection{Graph Components}

\begin{figure*}[tb]
    \centering
    \includegraphics[width=\linewidth]{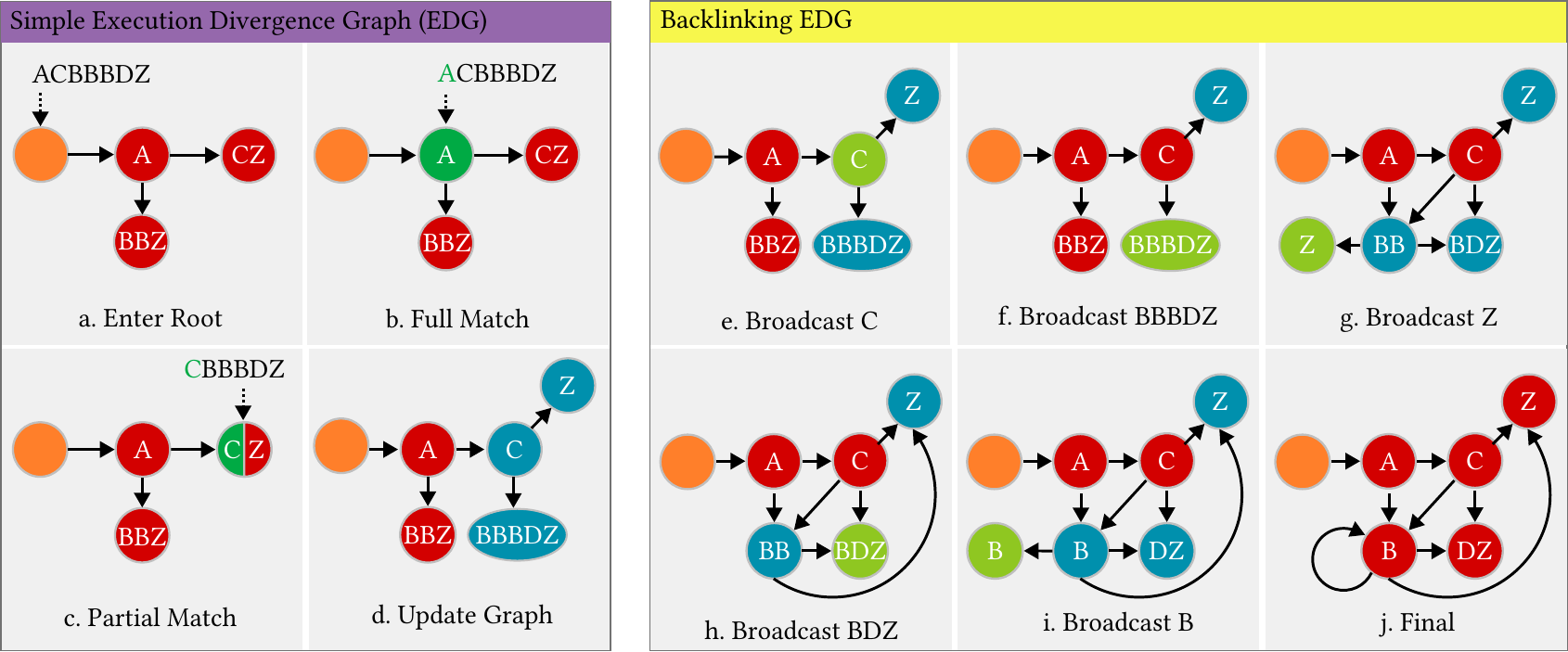}
    \caption{This illustration contrasts how the Simple and Backlinking Execution Divergence Graphs (EDGs) integrate the new trace \texttt{ACBBBDZ}: when a partial match is found (node \texttt{CZ}, step c), the Simple EDG immediately appends the remainder of the trace (step d), while the Backlinking EDG broadcasts all new segments (\texttt{C}, \texttt{Z}, \texttt{BBBDZ}) across the graph to check for identical segments. The orange node is the root, dark green nodes signify a full match, half dark-green/red nodes (e.g., \texttt{C/Z}) denote a partial match, and blue nodes represent newly created nodes; the light green nodes in Backlinking EDG indicate the node is currently broadcast.}
    \label{fig:edg_demo}
\end{figure*}

In the EDG, nodes represent execution segments discovered during the target’s runtime (except for the root node), and directed edges connect these segments to describe transitions between execution segments. Because the edges are directed, an edge from Node \texttt{X} to Node \texttt{Y} implies a parent--child relation, meaning that during execution the target executes \texttt{X} first and then proceeds to \texttt{Y}. As shown in step (a) of Figure~\ref{fig:edg_demo}, there are three non-root nodes corresponding to the execution segments \texttt{A}, \texttt{BBZ}, and \texttt{CZ}. Starting from the root, the only outgoing edge leads to Node \texttt{A}, indicating that \texttt{A} is the first execution segment observed at runtime. Node \texttt{A} has two outgoing edges to its children, Node \texttt{BBZ} and Node \texttt{CZ}, illustrating the two possible execution paths that follow segment \texttt{A}. Both \texttt{BBZ} and \texttt{CZ} have no outgoing edges, meaning they represent terminal segments. By following the edges starting from the root, the full set of execution paths occurring during runtime can be reconstructed.

\subsection{Match First Traversal}\label{subsec:match_first_traversal}

Our \textit{Match First Traversal} algorithm defines how an incoming trace traverses the simple EDG. At each step, the algorithm evaluates all adjacent nodes to determine the next node to explore based on the length of the common prefix. For example, when the trace \texttt{ACBBBDZ} is processed by the EDG (Figure~\ref{fig:edg_demo}, step a), it begins at the root and matches Node \texttt{A}. Because Node \texttt{A} fully matches the first part of the trace, this is a full match, meaning that Node \texttt{A} becomes the next node to be explored and the graph remains unchanged (step b). After the match, the algorithm consumes \texttt{A} from the trace, resulting in \texttt{CBBBDZ}, and proceeds to the next step. In step c, the algorithm finds that \texttt{CBBBDZ} and Node \texttt{CZ} share a common \texttt{C} but diverge afterward. This partial match causes the EDG to factor out the shared segment \texttt{C} as a new node, with the remaining segment \texttt{Z} becoming another new node connected as a child of \texttt{C}. The unmatched rest of the trace, \texttt{BBBDZ}, also diverges after \texttt{C} and is therefore added as another child of the new node \texttt{C} (step d).

\subsection{Backlinking Candidates}\label{subsec:backlink_candidates}

While the simple EDG directly updates the graph by adding the discovered new segments \texttt{C}, \texttt{Z}, and \texttt{BBBDZ} (step d in Figure~\ref{fig:edg_demo}), the Backlinking EDG instead broadcasts these new segments across the entire graph for additional matching. As shown in step e, the graph first broadcasts Node \texttt{C} but finds no match. It then continues with Node \texttt{BBBDZ} and discovers that Nodes \texttt{BBZ} and \texttt{BBBDZ} share the common trace segment \texttt{BB} (step f). The graph then factors out this shared segment as a new node \texttt{BB}, which is connected to two new nodes corresponding to the divergent parts \texttt{BDZ} and \texttt{Z}. Node \texttt{Z}, which originates from Node \texttt{BBZ}, is broadcast next, as shown in step g, where the graph finds that it perfectly matches another Node \texttt{Z}. These two Nodes \texttt{Z} are then merged into a single Node \texttt{Z}, which inherits the parents \texttt{BB} and Node \texttt{C} (step h). After broadcasting Node \texttt{Z}, the graph broadcasts Node \texttt{BDZ} and discovers that Node \texttt{BDZ} and Node \texttt{BB} share the common trace segment \texttt{B}. Consequently, a new Node \texttt{B} is created and connected to two new children, Node \texttt{DZ} and Node \texttt{B}. The graph then applies backlinking to the new child Node \texttt{B} in step i and finds a node with an identical trace segment, causing the two Nodes \texttt{B} to merge. Because these two Nodes \texttt{B} were connected by an edge before merging, the graph creates a \textit{self-loop edge} for the new Node \texttt{B}. Once step i is completed, three nodes remain in the broadcast queue: \texttt{B}, \texttt{Z}, and \texttt{DZ}; however, none of them matches any other node in the graph, so no further updates occur and the backlinking process ends (step j). Comparing the results of the simple EDG (step d) and the backlinking EDG (step j), the backlinking version yields fewer redundant trace segments in the final graph.

\section{Case Study}
\label{sec:case_study}

\begin{figure}[ptbh]
    \centering
    \begin{subfigure}{\linewidth}
        \centering
        \includegraphics[width=0.8\linewidth]{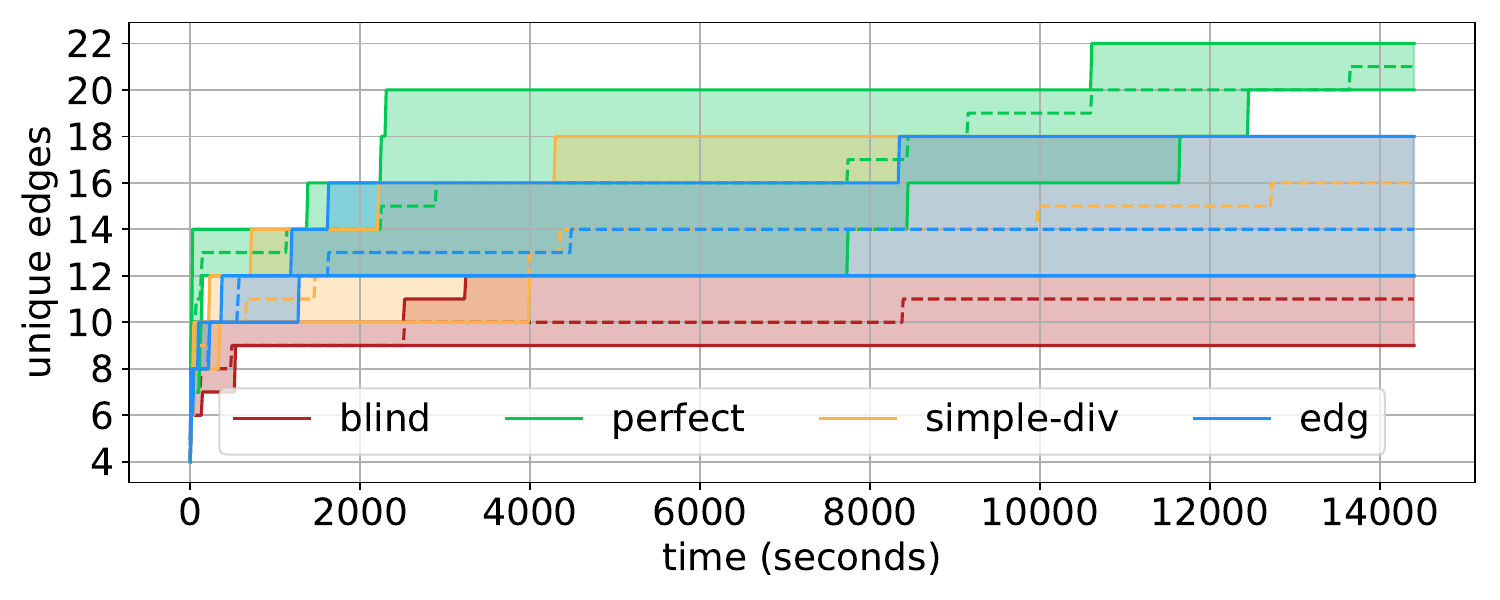}
        \caption{Password checker (Case 1).}
        \label{fig:pwd-check-4h}
    \end{subfigure}
    \begin{subfigure}{\linewidth}
        \centering
        \includegraphics[width=0.8\linewidth]{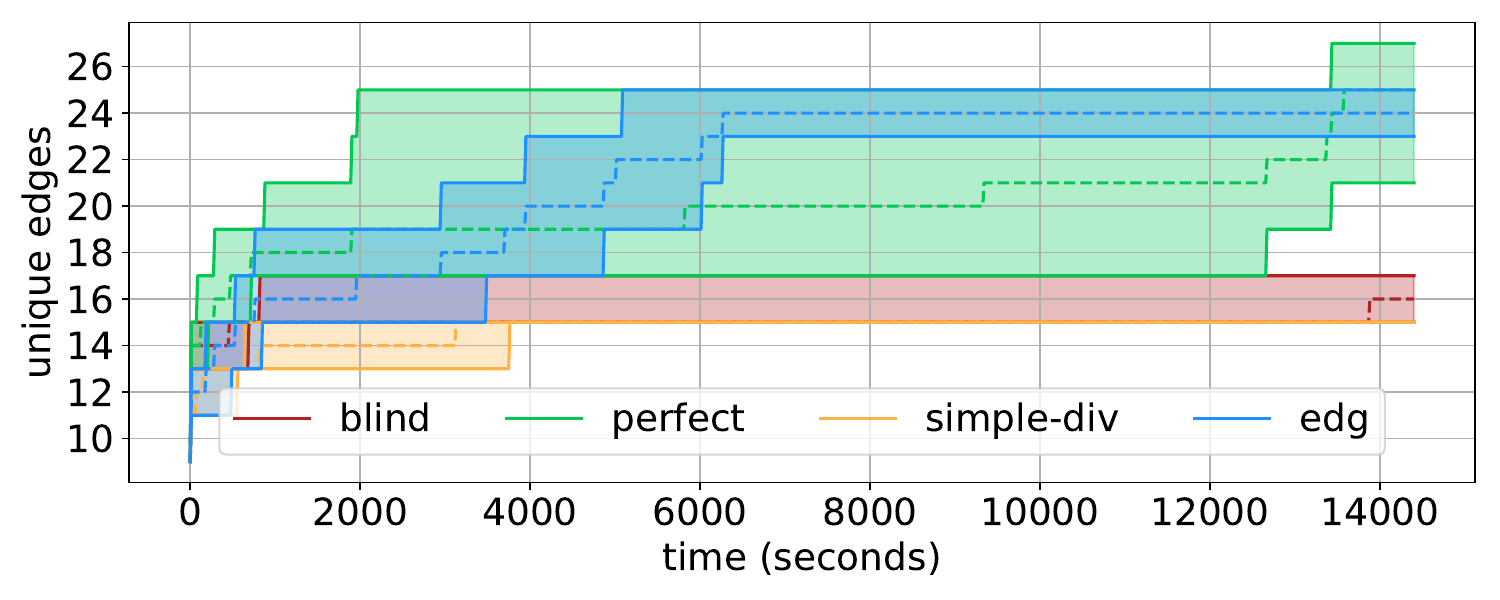}
        \caption{Password checker with input validator (Case 2).}
        \label{fig:pwd-input-check-4h}
    \end{subfigure}
    \begin{subfigure}{\linewidth}
        \centering
        \includegraphics[width=0.8\linewidth]{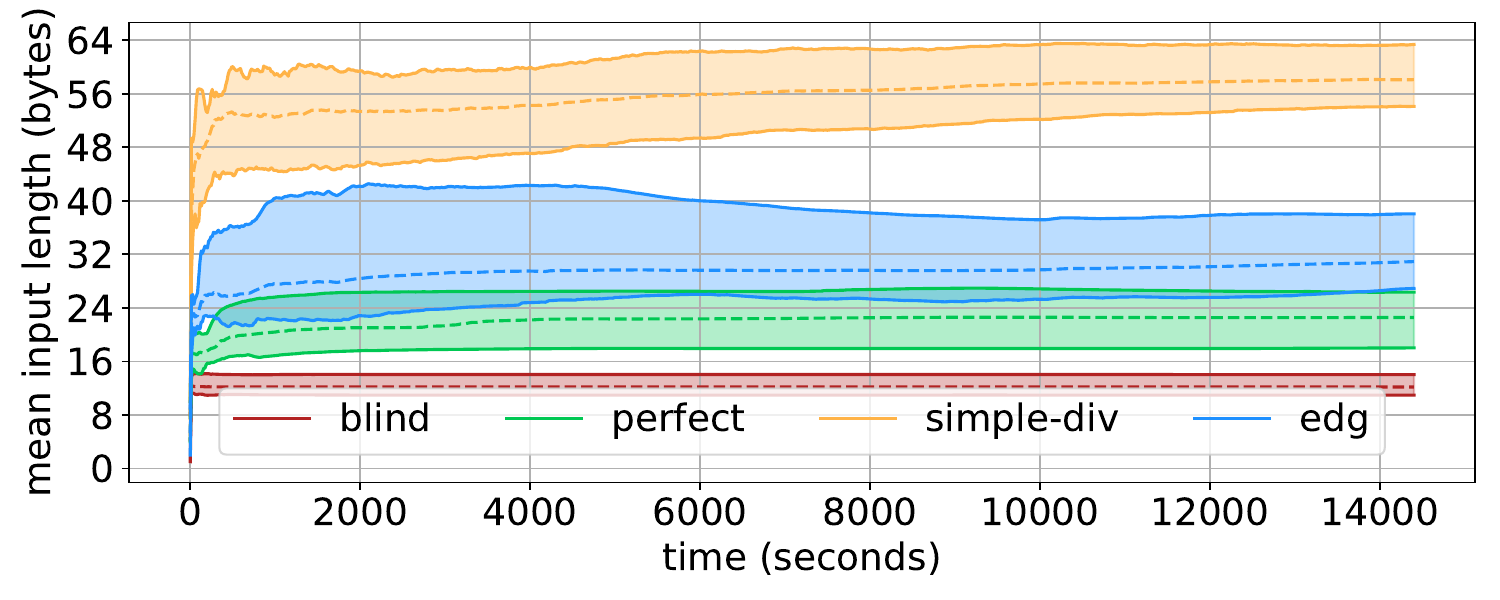}
        \caption{Average length of inputs over time (Case 2).}
        \label{fig:nestif-avg-inputlen-input-check-4h}
    \end{subfigure}
    \caption{Figure~\ref{fig:pwd-check-4h} and Figure~\ref{fig:pwd-input-check-4h} demonstrate the number of unique edges discovered over time for the two password checkers across four fuzzing scenarios for 4 hours: \emph{blind}, \emph{perfect}, \emph{simple-div}, and \emph{EDG}. The feedback provided by the divergence-detection-based approaches, \emph{simple-div} and \emph{EDG}, can effectively guide the fuzzer, since both outperform the \emph{blind} scenario, in which the fuzzer receives no feedback (Figure~\ref{fig:pwd-check-4h}). However, \emph{simple-div} performs poorly in Figure~\ref{fig:pwd-input-check-4h} because of repeated basic blocks introduced by the input validator, namely the loop that checks input bytes one by one for null values. These repeated basic blocks are mistakenly treated as divergences and mislead the fuzzer into generating longer inputs instead of focusing on interesting bytes related to the password (Figure~\ref{fig:nestif-avg-inputlen-input-check-4h}).}
    \label{fig:case-study-4h}
\end{figure}

In this section, we first use a simple password checker program as a demonstration of how different levels of feedback affect fuzzing performance. We show how the simple divergence approach easily generates a high number of false positives when the password checker includes an input validator. Hence, we introduce the Execution Divergence Graph to solve this problem. Lastly, we fuzz obfuscated password checkers to demonstrate how obfuscation affects the feedback available to the fuzzer.

\subsection{Fuzzing Scenarios}~\label{subsec:fuzz_scenarios_in_sec5}
We consider four main fuzzing scenarios: \emph{blind}, \emph{perfect}, \emph{simple-div}, and \emph{EDG}. In the \emph{blind} scenario, the fuzzer receives no feedback from the target. This is the baseline scenario when the target cannot be instrumented. The \emph{perfect} scenario provides ground-truth edge information to the fuzzer. We use this setting as a reference for fuzzer performance under ideal conditions. The \emph{simple-div} (Section~\ref{subsec:simplediv}) and \emph{EDG} (Section~\ref{subsec:edg}) scenarios provide feedback derived from execution traces. For a fuzzer guided by the \emph{simple-div} mechanism, every trace that has not been seen before is treated as a new discovery. In contrast, in the \emph{EDG}-based approach, all collected traces are used to reconstruct a graph that provides more fine-grained trace information, indicating which segments of an incoming trace have already been observed and which are new.

\begin{algorithm}
	\SetAlgoLined
	\KwData{$data\_ptr$: input buffer}
	\If{$data\_ptr[0] = \texttt{'b'}$}{
		\If{$data\_ptr[1] = \texttt{'a'}$}{
			\If{$data\_ptr[2] = \texttt{'d'}$}{
				\If{$data\_ptr[3] = \texttt{'f'}$}{
					\If{$data\_ptr[4] = \texttt{'u'}$}{
						\If{$data\_ptr[5] = \texttt{'z'}$}{
							\If{$data\_ptr[6] = \texttt{'z'}$}{
								\If{$data\_ptr[7] = \texttt{'!'}$}{
									\textbf{accept password}\;
								}
							}
						}
					}
				}
			}
		}
	}
	\Return $0$\;
	\caption{Pseudo code of the password checker.}
	\label{alg:password-checker}
\end{algorithm}

\subsection{Password Checker}\label{subsec:pwd_check}

We implement a password checker with secret \texttt{badfuzz!} (Alg~\ref{alg:password-checker}) and run an AFL fuzzer across the four scenarios. Under each scenario, we repeat the fuzzing process five times. Each fuzzing run lasts 4 hours because the fuzzer under the \emph{perfect} scenario already recovers the password. As Figure~\ref{fig:pwd-check-4h} shows, the fuzzer under the \emph{perfect} scenario discovers the highest number of unique edges (22). The fuzzer without feedback (the \emph{blind} scenario) performs the worst among all scenarios and finds only 12 edges in the best case. When the fuzzer uses trace segments as feedback (\emph{simple-div} and \emph{EDG}), it performs better than in the no-feedback setting. Since all execution paths are related to the characters in the password, the new trace segments found by \emph{simple-div} and \emph{EDG} are identical. Hence, the fuzzer shows comparable performance under these two scenarios.

\subsection{Password Checker + Input Validator}\label{subsec:pwd_check_input}

However, as we discussed in Section~\ref{subsec:simplediv}, the simple divergence approach only detects whether the full trace has been observed before. Any small difference, e.g., if a code segment repeats more times than in previous traces, causes the simple divergence approach to classify the trace as novel, which is a false positive. To demonstrate this issue, we add an input validator to the previous target. The input validator processes the input string and checks each byte for a null value. When the program encounters a null value, it terminates immediately. 

Due to this construction, the number of iterations through the input validator depends on the content and length of the input. This causes traces generated from different input lengths to be regarded as novel by the simple divergence approach. As expected, we observe a substantial performance impact in the \textit{simple-div} scenario, as Figure~\ref{fig:pwd-input-check-4h} shows: the fuzzer discovers only a very limited number of unique edges. In addition, \emph{simple-div} leads the fuzzer to generate longer inputs than in other scenarios (Figure~\ref{fig:nestif-avg-inputlen-input-check-4h}) because of these input-length-related false positives. As a result, the password checker takes more time to process long inputs than short ones. Because of the longer processing time, the fuzzer generates fewer inputs during a fixed fuzzing duration (\emph{EDG}: ~1.6M inputs vs \emph{simple-div}: ~1.2M inputs). The \emph{EDG} scenario, on the other hand, is still able to reach deeper parts of the program. These results show that the backlinking mechanism in \emph{EDG} (Section~\ref{subsec:edg}) provides a substantial improvement in fuzzing performance.

\subsection{Obfuscated Password Checker}

\begin{figure*}[t]
    \centering
    \begin{subfigure}[t]{0.63\linewidth}
        \centering
        \includegraphics[width=\linewidth]{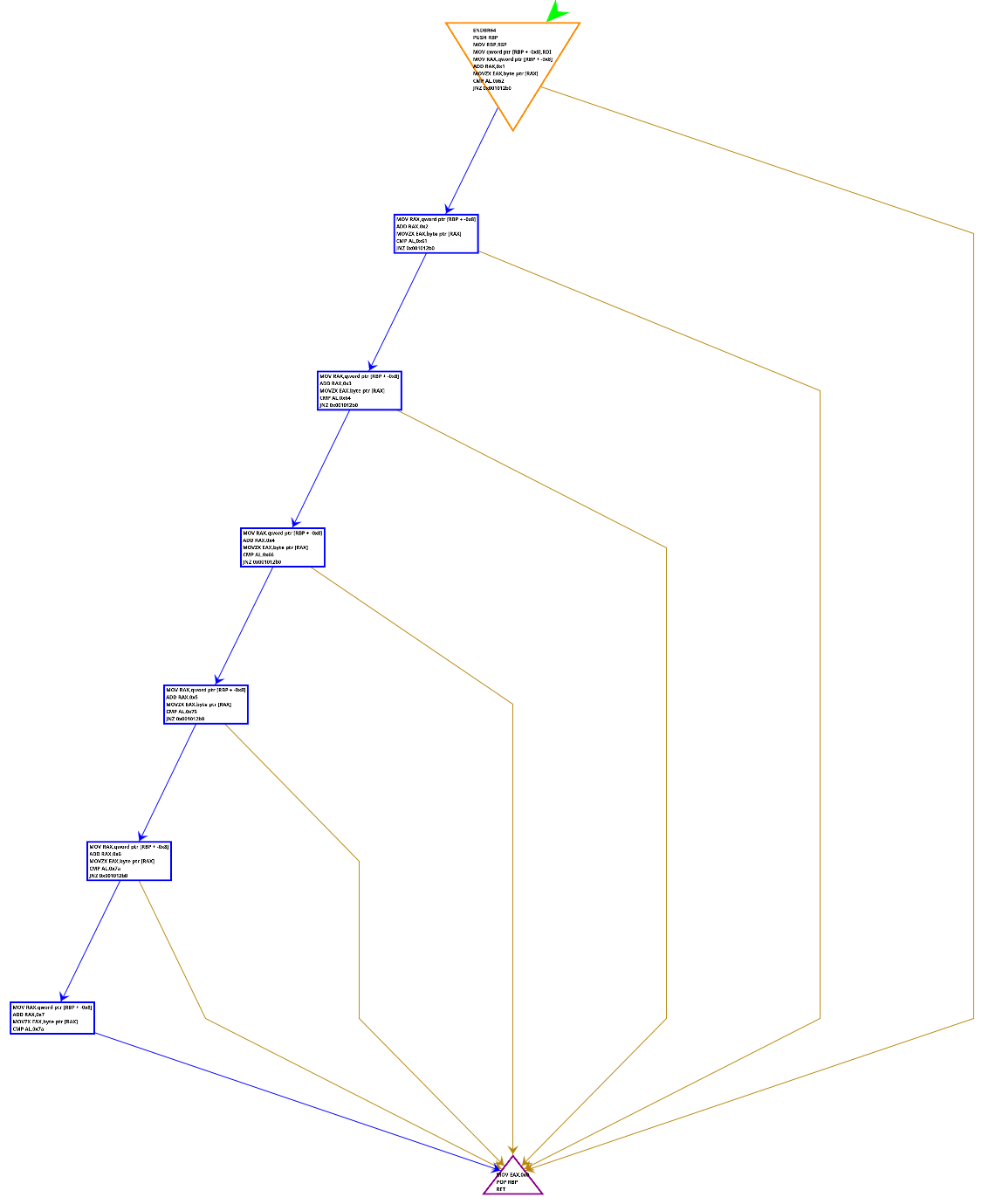}
        \caption{The CFG without obfuscation.}
        \label{fig:cfg-pwd-check}
    \end{subfigure}
    \hfill
    \begin{subfigure}[t]{0.36\linewidth}
        \centering
        \includegraphics[width=\linewidth]{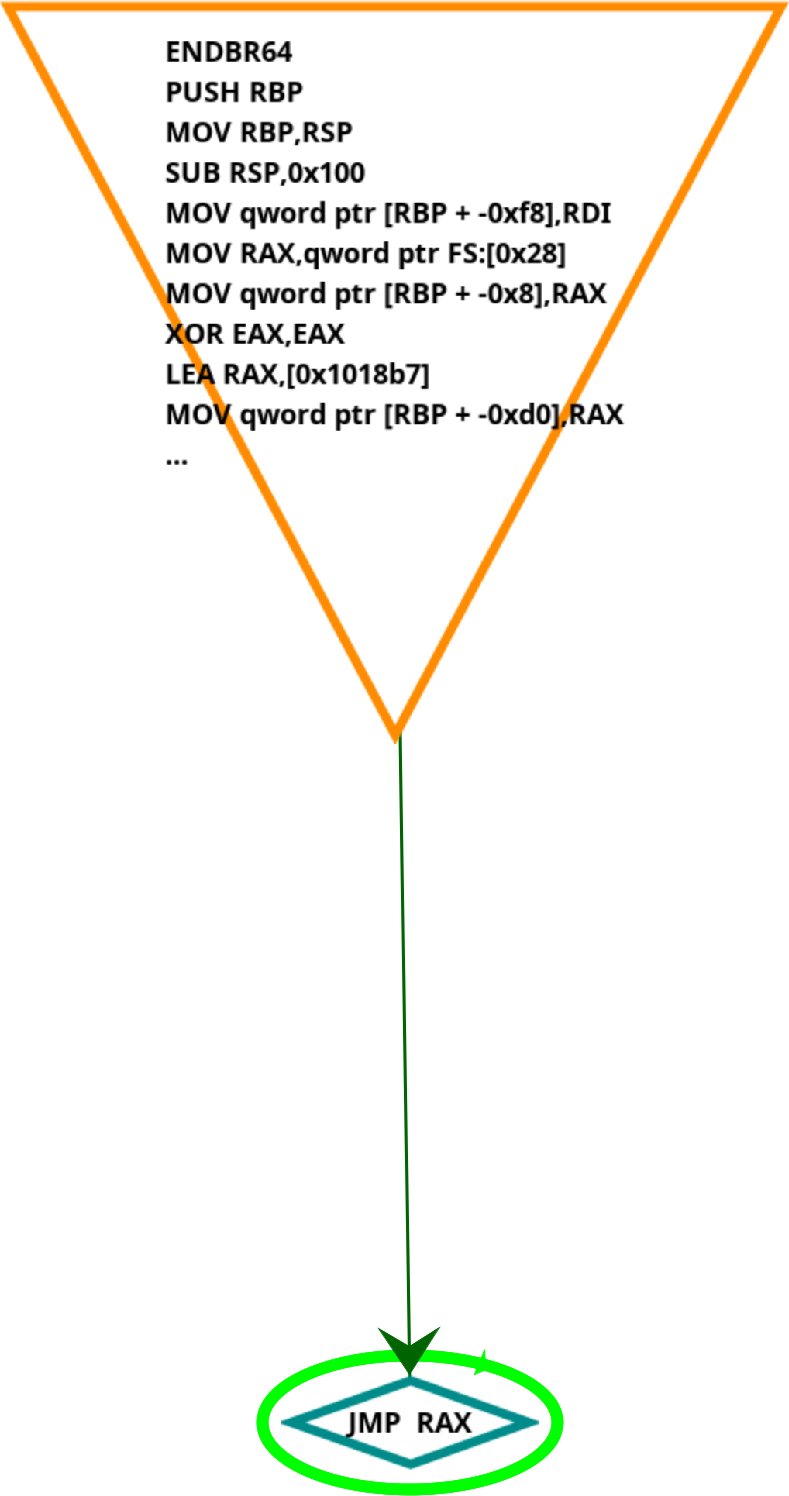}
        \caption{The CFG with obfuscation.}
        \label{fig:cfg-pwd-check-indirect}
    \end{subfigure}
    \caption{The comparison between the CFGs of the normal and the obfuscated binaries. The binary is the password checker mentioned in Section~\ref{subsec:pwd_check}. Before the obfuscation, the CFG shows explicit links between basic blocks and the true/false paths (Figure~\ref{fig:cfg-pwd-check}). The obfuscation turns the branch instructions into jump instructions, whose destination addresses can only be obtained after execution. Therefore, these jump addresses cannot be found by static instrumentation tool before the program starts (\texttt{JMP RAX} in Figure~\ref{fig:cfg-pwd-check-indirect}).}
    \label{fig:cfg-pwd-check-overview}
\end{figure*}

\begin{figure*}[t]
    \centering
    \begin{subfigure}[t]{0.49\linewidth}
        \centering
        \includegraphics[width=\linewidth]{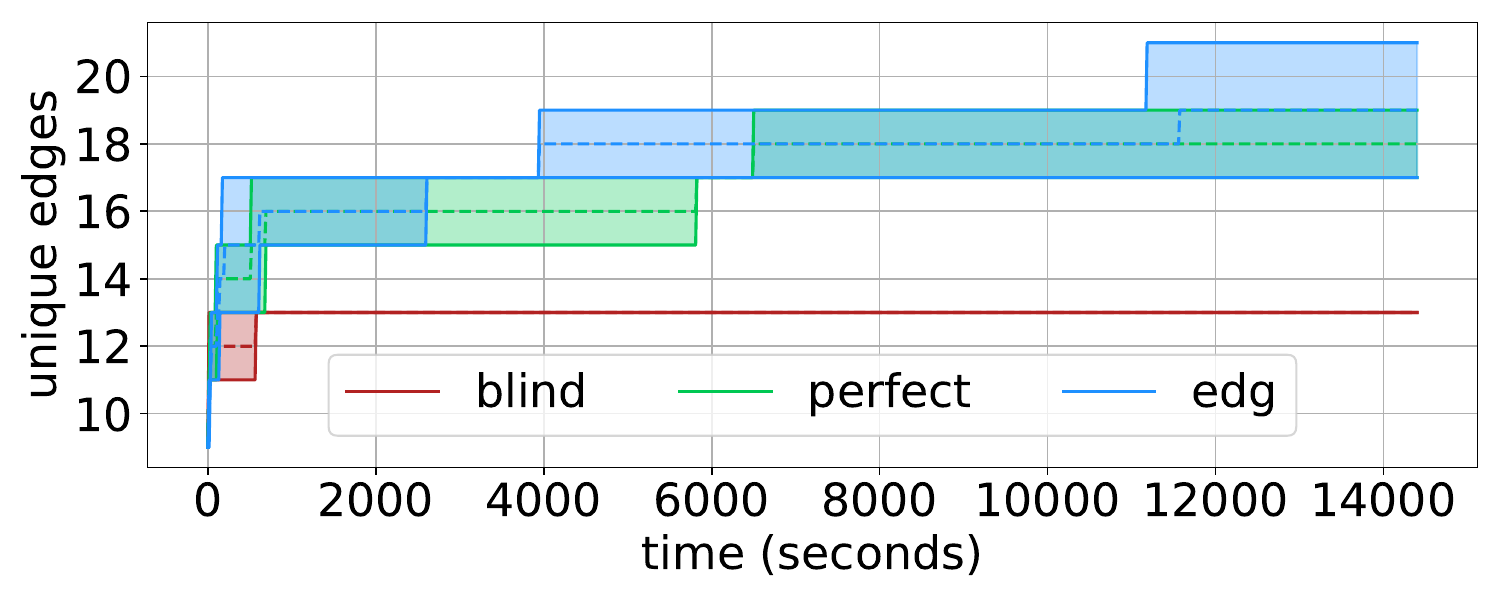}
        \caption{Obfuscated with Control-flow Flatten.}
        \label{fig:fuzz-flatten-pwd-check}
    \end{subfigure}
    \hfill
    \begin{subfigure}[t]{0.49\linewidth}
        \centering
        \includegraphics[width=\linewidth]{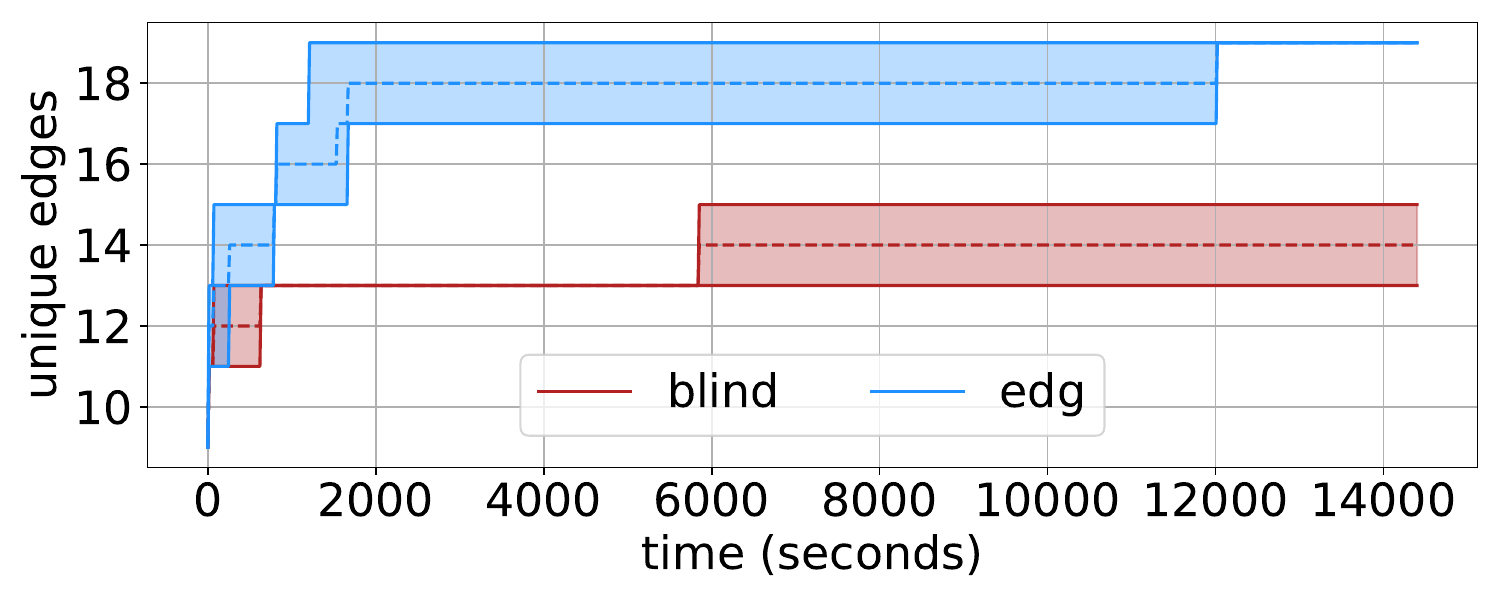}
        \caption{Obfuscated with Indirect Jump.}
        \label{fig:fuzz-indirect-pwd-check}
    \end{subfigure}
    \caption{The number of edges discovered while fuzzing the password checker under two different obfuscation schemes is shown for control-flow flattening (Figure~\ref{fig:fuzz-flatten-pwd-check}) and indirect jumps (Figure~\ref{fig:fuzz-indirect-pwd-check}). We obtain these edge counts from the non-obfuscated binary: we first fuzz the obfuscated binaries and then trace the execution of the non-obfuscated binary using the inputs generated during fuzzing. Because both obfuscation schemes significantly modify the original CFG, we observe degraded fuzzing performance compared with Figure~\ref{fig:pwd-check-4h} and Figure~\ref{fig:pwd-input-check-4h}. However, \emph{EDG} and \emph{perfect} still outperform the \emph{blind} fuzzer in the control-flow-flattening case. In contrast, \emph{perfect} no longer works in the indirect-jump case because the control-flow targets are resolved only during execution and cannot be determined at compile time. In this case, the \emph{perfect} fuzzer is effectively reduced to a \emph{blind} fuzzer. Given access to runtime execution traces, \emph{EDG}, in contrast, can still guide the fuzzer by identifying divergences between executions.}
    \label{fig:fuzz-obf-pwd-check}
\end{figure*}

We discuss the obfuscated case in which \emph{EDG} is a suitable replacement for conventional edge feedback for an AFL fuzzer, particularly in cases where static instrumentation is infeasible.

\parhead{Password Checker's CFG} The CFG of the non-obfuscated password checker is shown in Figure~\ref{fig:cfg-pwd-check}. The figure shows that the password checker has a nested-\texttt{if} structure, in which the execution of each basic block depends on the execution of the previous one, except for the entry and exit points.

\parhead{Control-flow Flattening} Control-flow flattening~\cite{wang2001security} is one of the most common techniques for obfuscating binaries. It transforms common control structures, such as \texttt{if} statements, loops, and function calls, into a central dispatcher that determines which basic block to execute based on its current state. Through control-flow flattening, the password checker’s nested-\texttt{if} structure is completely flattened and transformed into multiple cases within a \texttt{switch} structure. For example, when the non-obfuscated password checker processes the input \texttt{bad}, it directly traverses the first \texttt{if} statement (\texttt{data\_ptr[0] == 'b'}) up to the third one (\texttt{data\_ptr[2] == 'd'}). In contrast, the control-flow-flattened version first enters the dispatcher and is directed to the case corresponding to \texttt{data\_ptr[0] == 'b'}. The program then updates the dispatcher’s state and returns to the dispatcher again. Based on the updated state, the dispatcher transfers execution to the next case, \texttt{data\_ptr[1] == 'a'}. This example shows that additional dispatcher blocks are introduced into the obfuscated binary. These extra blocks complicate fuzzing and therefore degrade performance in the 4-hour fuzzing experiment (Figure~\ref{fig:fuzz-flatten-pwd-check}) compared with the non-obfuscated binary (Figure~\ref{fig:pwd-input-check-4h}). The mean number of discovered unique edges decreases from 24.75 to 18 for \emph{perfect} and to 19 for \emph{EDG}. With longer fuzzing time, the fuzzer can still recover the whole password, since it can still obtain edge feedback from the instrumented binary.

\parhead{Indirect Jump} To make the code more obfuscated and harder to instrument automatically, the obfuscator leverages \emph{indirect jump}~\cite{LinnD03} to replace branch instructions. Instead of transferring execution to an explicit destination (e.g., the address of a basic block), an indirect jump computes the destination at runtime, typically from a register or a memory slot. As a result, the successor basic block is no longer explicit in the instruction itself, which obscures the control-flow graph. In practice, the obfuscator first splits the original code into multiple basic blocks, assigns each block a corresponding entry in a jump table, and then uses a state variable to select the next target dynamically. Because the branch target is resolved only during execution, analysis and static instrumentation tools have a harder time recovering the original execution order and accurately identifying branch boundaries. Without instrumentation, the original coverage-guided fuzzer becomes blind, since it cannot obtain any edge feedback during fuzzing. However, \emph{EDG} can still guide the fuzzer because it detects changes in execution by comparing traces at runtime. Figure~\ref{fig:fuzz-indirect-pwd-check} demonstrates that the fuzzer guided by \emph{EDG} feedback reaches deeper parts of the program than the unguided fuzzer.

\section{Evaluation}

In this section, we first evaluate how efficiently our divergence-detection-based approaches can guide the fuzzer when testing non-obfuscated binaries (Section~\ref{subsec:guide_fuzzer_inst_traces_div}). Next, we obfuscate some binaries to hide their control flow in a jump table, which cannot be statically instrumented. We then evaluate whether our approaches can still provide edge feedback that enables the fuzzer to test the obfuscated binaries (Section~\ref{subsec:fuzz_obf_indirect_jump}).

\subsection{Settings}

\parhead{Fuzzer} We use LibAFL~\cite{libafl} for fuzzing our non-obfuscated and obfuscated binaries. The number of edges in \texttt{EDGE\_MAP} is 262,144, which defines the maximum number of edges that can have their own unique IDs. If a binary has more than 262,144 edges, some edges will share the same ID. In terms of feedback, we use \texttt{MaxMapPow2Feedback} to enable hit-count buckets. We choose \texttt{PowerSchedule}~\cite{BohmePR16} from the default schedulers, since the scheduler prioritizes inputs that trigger uncommon edges. Hence, the scheduler encourages the fuzzer to explore more untouched code in the binaries.

\parhead{Instruction Tracing} We leverage DynamoRIO~\cite{dynamorio} to trace instructions during runtime and store the traced instructions in shared memory, where our divergence-detection-based approaches can access them.

\parhead{Obfuscation} Tigress~\cite{tigress} is a source-to-source obfuscator that provides various obfuscation options for researchers. We use it to convert C code into obfuscated code and compile the result with GCC.

\parhead{Scenarios} As in Section~\ref{subsec:fuzz_scenarios_in_sec5}, we evaluate three fuzzing scenarios: \emph{blind}, \emph{perfect}, and \emph{EDG}. In Section~\ref{subsec:guide_fuzzer_inst_traces_div}, we use the \emph{blind} scenario as the worst-case setting, in which the fuzzer receives no internal feedback. In Section~\ref{subsec:fuzz_obf_indirect_jump}, the \emph{blind} scenario represents the case in which static instrumentation fails because of the indirect-jump implementation in the obfuscated binaries. We evaluate the \emph{perfect} scenario only in Section~\ref{subsec:guide_fuzzer_inst_traces_div}, because the fuzzer cannot access basic-block information without static instrumentation. Hence, in Section~\ref{subsec:fuzz_obf_indirect_jump}, it is replaced by the \emph{blind} scenario. The \emph{EDG} scenario is our divergence-detection-based fuzzing setting, which has no false-positive problem compared to the \emph{simple-div} approach (Section~\ref{subsec:pwd_check_input}). 

\subsection{Guide Fuzzer with Instruction Trace Divergences}\label{subsec:guide_fuzzer_inst_traces_div}

\begin{figure*}[t]
    \centering
    \begin{subfigure}[t]{0.49\linewidth}
        \centering
        \includegraphics[width=\linewidth]{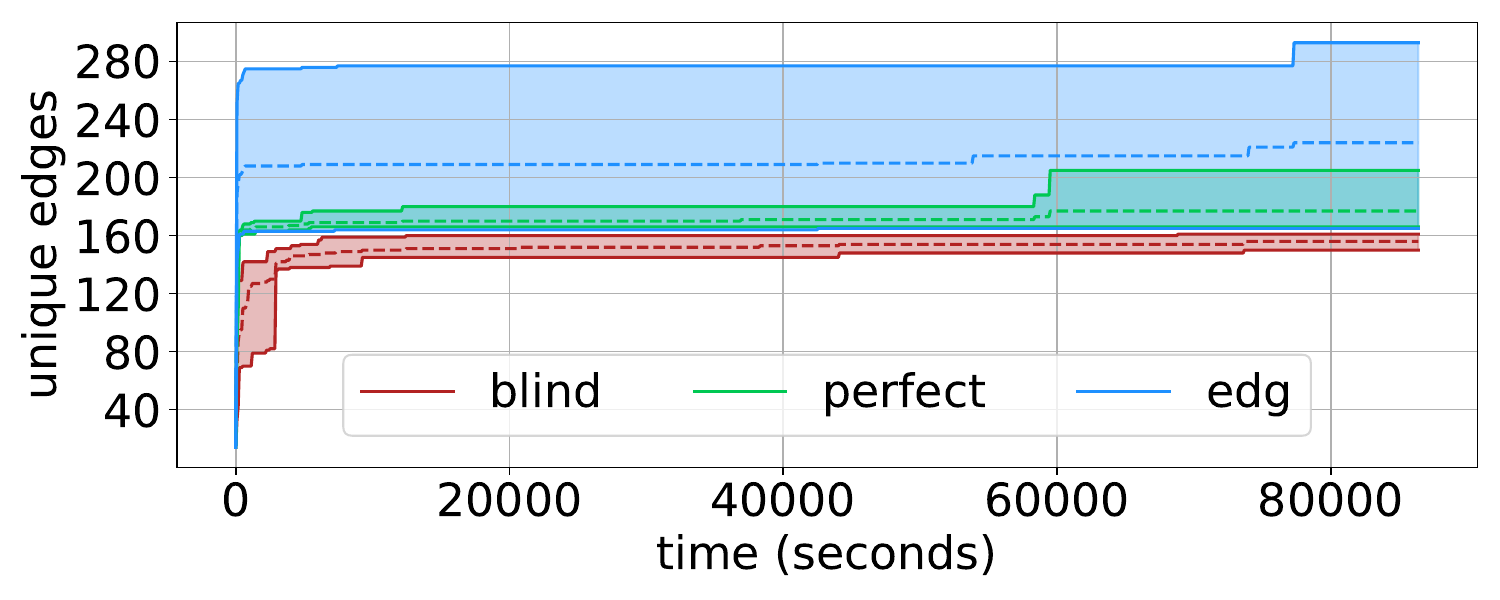}
        \caption{Results for minmea.}
        \label{fig:minmea-24h}
    \end{subfigure}
    \hfill
    \begin{subfigure}[t]{0.49\linewidth}
        \centering
        \includegraphics[width=\linewidth]{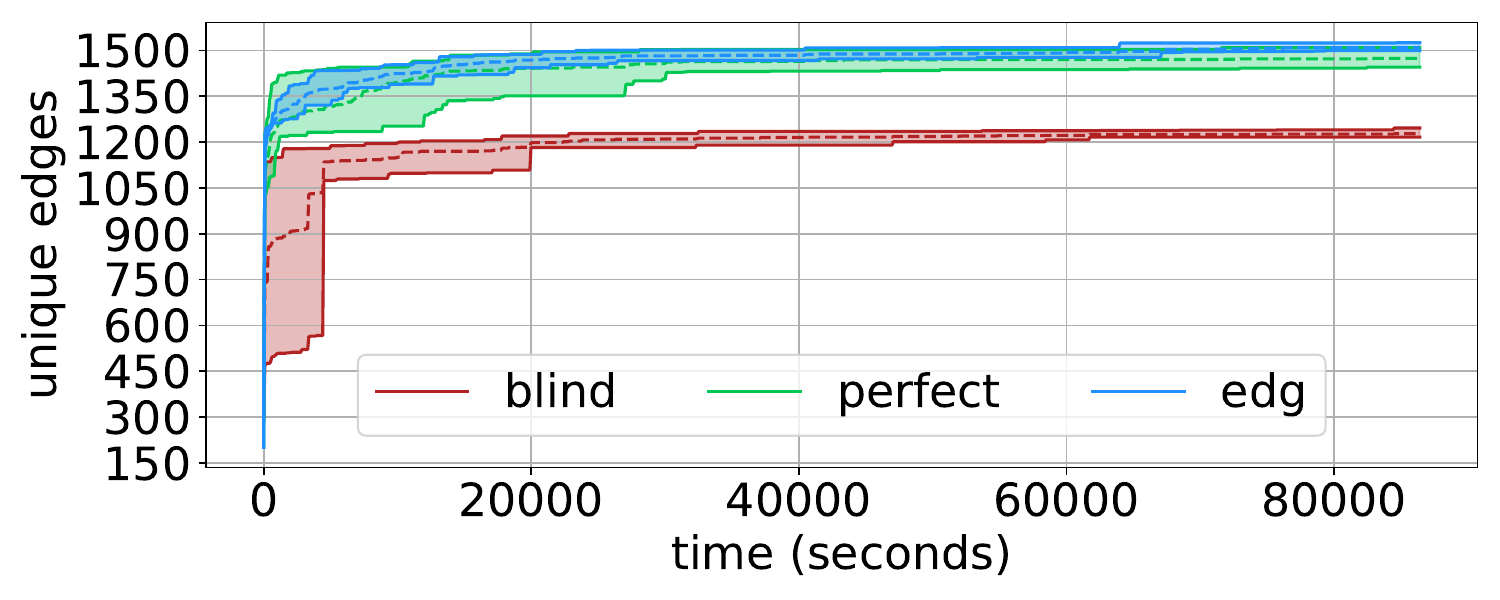}
        \caption{Results for libpng.}
        \label{fig:libpng-24h}
    \end{subfigure}
    \caption{Fuzzing results after 24 hours for fuzzers under three scenarios \emph{blind}, \emph{perfect}, and \emph{EDG}. The dashed line represents the mean number of discovered edges across five fuzzing runs, while the other two solid lines represent the maximum and minimum values, respectively. The results demonstrate that the divergence-detection-based approach (\emph{EDG}) achieves performance comparable to that of the fuzzer with ground-truth edge feedback (\emph{perfect}). More details are provided in Sec.~\ref{subsec:guide_fuzzer_inst_traces_div}.}
    \label{fig:guided_fuzzer_with_inst_div_24h}
\end{figure*}

We select two open-source libraries, minmea~\cite{minmea} and libpng~\cite{libpng}, to test whether execution divergences in these programs can guide the fuzzer. Each fuzzing run lasts 24 hours, and each scenario is repeated five times. The results for \emph{the number of discovered unique edges over time} are shown in Figure~\ref{fig:guided_fuzzer_with_inst_div_24h}.

\parhead{minmea} minmea is a GPS parser that supports multiple sentence formats in the NMEA 0183 standard~\cite{nmea0183wiki}. It is specifically designed for embedded systems because it is implemented in ISO C99 and has no dynamic memory allocation. After the fuzzing experiment, we find that our \emph{EDG}-based fuzzing outperforms the other scenarios. \emph{EDG} discovers 224.4 edges on average, whereas \emph{perfect} and \emph{blind} discover 176.6, and 155.6 edges, respectively (Figure~\ref{fig:minmea-24h}). 
The main reason \emph{EDG} discovers more edges than \emph{perfect} is that EDG generates inputs
 that cause segmentation faults (157 inputs). These inputs trigger system calls in the OS for crash handling, and thus the basic-block transitions within those system calls are recorded among the discovered edges.

\parhead{libpng} Libpng is a widely used PNG parser library. We use Libpng version 1.6.37 provided in LibAFL's fuzzing example. The fuzzing result shows a similar outcome to minmea (Figure~\ref{fig:libpng-24h}): \emph{EDG}-assisted fuzzing reaches the highest average number of edges (1509.6) among all scenarios. The remaining scenarios, from second high to low, are \emph{perfect} (1473.8) and \emph{blind} (1227.4). In addition, \emph{EDG} also generates inputs (140) that cause segmentation faults.

In summary, we show that the fuzzer can leverage execution divergences to guide fuzzing and performs comparably to a fuzzer with full edge feedback. Whether these inputs reveal genuine bugs or exception-handling deficiencies in the test programs remains an open question for future work.

\subsection{Fuzzing Indirect-Jump Binary}\label{subsec:fuzz_obf_indirect_jump}

\begin{figure*}[t]
    \centering
    \begin{subfigure}[t]{0.49\linewidth}
        \centering
        \includegraphics[width=\linewidth]{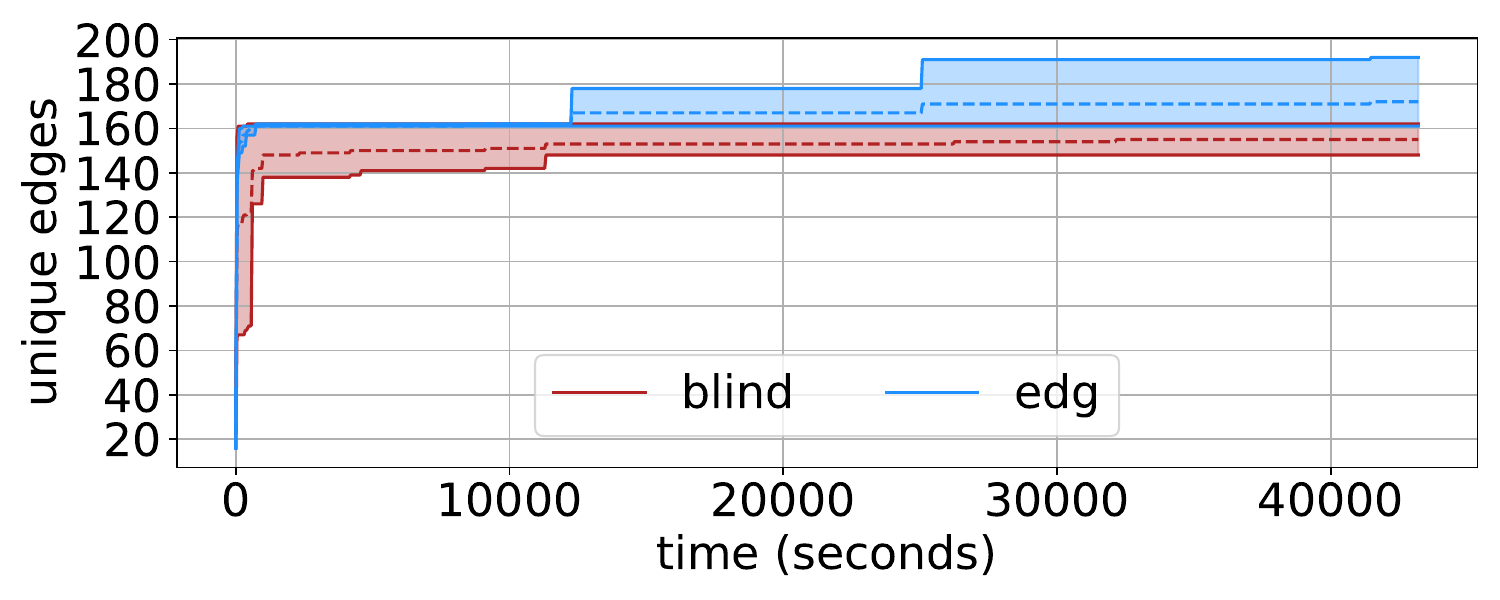}
        \caption{Obfuscated minmea.}
        \label{fig:obf-minmea-12h}
    \end{subfigure}
    \hfill
    \begin{subfigure}[t]{0.49\linewidth}
        \centering
        \includegraphics[width=\linewidth]{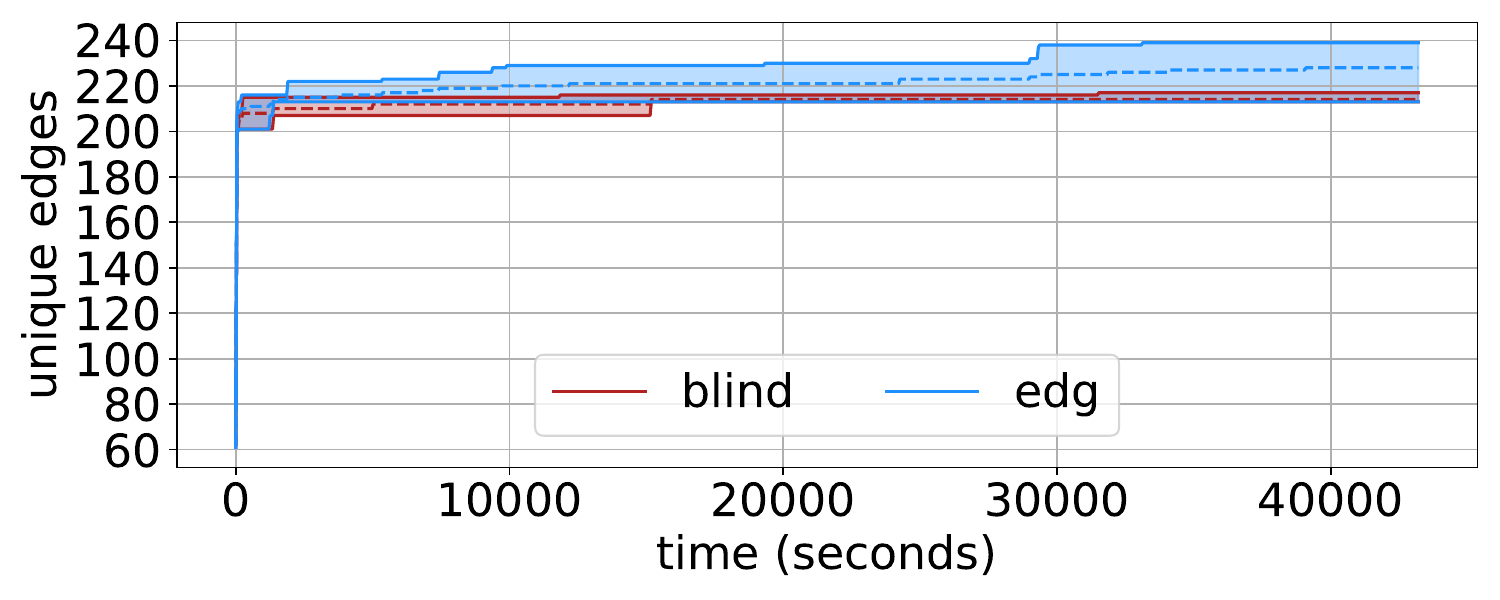}
        \caption{Obfuscated Lz4.}
        \label{fig:obf-lz4-12h}
    \end{subfigure}
    \caption{Results of fuzzing real-world binaries obfuscated with indirect jumps after 12 hours. Since both binaries use indirect jumps to hide control flow, the \emph{perfect} setting effectively reduces to the \emph{blind} setting. Without feedback, \emph{blind} performs worse than \emph{EDG}, which can still guide the black-box fuzzer when static instrumentation is impossible. More details are provided in Section~\ref{subsec:fuzz_obf_indirect_jump}.}
    \label{fig:fuzz_obf_binaries_12h}
\end{figure*}

We investigate whether our \emph{EDG}-based fuzzing can help the fuzzer test obfuscated binaries whose basic-block addresses are hidden through a jump table. Since the jump targets are resolved only during execution, static instrumentation cannot determine these addresses in the obfuscated binaries. As a result, the \emph{perfect} setting effectively reduces to \emph{blind} fuzzing. Because \emph{simple-div} causes memory-exhaustion problems when fuzzing obfuscated binaries as well, we focus only on \emph{EDG}, our optimized divergence-detection-based approach, in this section. Note that the number of discovered unique edges is measured from the \emph{non-obfuscated} version of the binary. We reuse the inputs generated while fuzzing the obfuscated binaries, replay them on the non-obfuscated binaries, and trace the edges. For each obfuscated program, we run fuzzing under the \emph{EDG} and \emph{blind} scenarios for 12 hours and repeat each scenario three times.

\parhead{Obfuscated Minmea} As Figure~\ref{fig:obf-minmea-12h} demonstrates, \emph{EDG} overall performs better than \emph{blind} in terms of the number of unique edges discovered in the non-obfuscated \texttt{minmea} binary (171.67 vs 155). Neither scenario triggers any segmentation fault observed in Section~\ref{subsec:guide_fuzzer_inst_traces_div}.

\parhead{Obfuscated lz4} lz4~\cite{lz4} is a well-known compression algorithm and is widely shipped in many operating systems. After the fuzzing evaluation, \emph{EDG} still performs better than \emph{blind} (228 vs 214.33, Figure~\ref{fig:obf-lz4-12h}), even though the blind fuzzer achieves 20\% higher throughput. However, \emph{EDG} in the worst case performs as poorly as the worst-case \emph{blind} scenario: only 213 unique edges are discovered.

As a result, we conclude that \emph{EDG} can help black-box fuzzing when static instrumentation is impossible.

\section{Discussion}

\parhead{Anti-Fuzzing Protected Binary} Anti-Fuzzing approaches~\cite{GulerAAH19,JungHSPLK19} install fake paths in a binary to mislead coverage-guided fuzzers into exploring traps, which waste the fuzzers' time and computational resources. Since these fake paths are input-sensitive, even minor input mutations made by the fuzzers can easily trigger new edges along these fake paths. The fuzzers regard these edges as interesting and therefore spend more effort exploring the fake parts of the binary. Unfortunately, an \emph{EDG} cannot effectively mitigate fake paths either because these fake paths can create various execution paths during runtime, which in turn create divergences that also mislead \emph{EDG}.

\parhead{Dynamic JIT Obfuscation} Some advanced obfuscation techniques~\cite{myjit,tigress} leverage JIT compilation to generate and obfuscate machine code at runtime. Under these techniques, the same function may produce different machine code each time it is invoked during execution. Because \emph{EDG} cannot analyze the semantics of dynamically generated machine code, it treats all dynamically generated code originating from the same function as separate execution instances. As a result, \emph{EDG} produces a large amount of false-positive feedback for the fuzzer.

\parhead{EDG vs Dynamic Instrumentation} Although dynamic instrumentation tools like DynamoRIO~\cite{dynamorio} can discover the basic blocks hidden in a jump table by resolving the memory address during runtime, directly using such tools to provide edge feedback for the fuzzer cannot solve fuzzing scenarios where an obfuscated binary contains millions of bogus basic blocks that are always executed. When these bogus basic blocks are executed, DynamoRIO generates a huge number of edges as feedback, thereby overwhelming the fuzzer's edge map. In contrast, \emph{EDG} only generates edge feedback when meaningful execution-trace differences appear. We conclude that \emph{EDG} can assist fuzzing based on dynamic instrumentation when testing obfuscated binaries that leverage jump tables to hide basic blocks and contain an extremely large number of bogus basic blocks.

\parhead{EDG with other types of execution traces} \emph{EDG} can also be applied to non-instruction traces if the analyst has a robust model for detecting their divergences. For instance, the analyst can collect power traces from IoT devices to train a divergence-detection model and use the model together with \emph{EDG} to generate edge feedback for guiding a fuzzer.

\section{Related Work}

Non-instrumented fuzzing treats the target largely as a black box and relies on input structure and externally observable states rather than compiler-inserted coverage instrumentation. Grammar-based fuzzing leverages the input structure of a target to mutate inputs without direct instrumentation. Havrikov et al.~\cite{havrikov2019covering} propose systematic coverage of grammar productions via \emph{k}-paths. This approach benefits non-instrumented fuzzing because it improves the quality and diversity of generated inputs. Olsthoorn et al.~\cite{olsthoorn2020structured} combine search-based testing with grammar-based fuzzing to generate highly structured inputs. \emph{EDG} can assist grammar fuzzers by providing additional execution-level information about a non-instrumented target.

Several prior works use externally observable states to guide fuzzing. DistFuzz~\cite{zou2025distfuzz} uses inter-node messages as feedback for black-box fuzzing without source instrumentation. Similarly, Snipuzz~\cite{feng2021snipuzz} infers message snippets from IoT responses to guide mutation. Kim et al.~\cite{kim2023intender} propose Intender, which uses intent-state transitions as external guidance for network fuzzing and exposes semantic bugs. de Ruiter et al.~\cite{deruiter2015tls} infer TLS protocol state machines via black-box testing and inspect them for spurious transitions. RESTler~\cite{atlidakis2019restler} generates REST API request sequences from specifications to explore cloud states and infer producer-consumer dependencies. Chen et al.~\cite{chen2018iotfuzzer} present IoTFuzzer, which leverages protocol states between mobile apps and IoT devices. DifFuzz~\cite{nilizadeh2019diffuzz} is a resource-guided fuzzer for detecting time- and space-based side-channel vulnerabilities. Although \emph{EDG} is not directly applicable to all of these prior scenarios, \emph{EDG} can still be beneficial for state-guided fuzzers that test a single target entity.

\section{Conclusion}\label{sec:conclusion}

In this work, we discuss approaches for collecting suitable feedback for fuzzing in scenarios where neither classic binary instrumentation nor static analysis are possible. In particular, we aim to detect the execution of novel code segments. To differentiate execution traces, we introduce the concept of execution divergence and show how it can be used to detect novel code execution, which can subsequently guide the fuzzer. We first introduce the simple divergence detection approach (Section~\ref{subsec:simplediv}), which directly compares execution traces to determine whether the current execution trace is unique. We then argue that, while the simple approach can be used to guide a fuzzer, it struggles with repeated execution of code segments. To address this false-positive problem, we develop algorithms such as Match First Traversal and Backlinking Candidates to reconstruct a graph from execution traces, namely the \emph{Execution Divergence Graph} (Section~\ref{sec:edg_in_detail}). This EDG is essentially a CFG-like structure based on the code observed during execution so far.

We implement the approach and experimentally demonstrate that execution-divergence-based approaches are effective for fuzzing real-world programs. We show that, in many scenarios, our EDG-based approach reaches similar coverage to that of traditional full instrumentation of the target, assuming such instrumentation is possible. Our results also show that the EDG approach enables feedback in scenarios where traditional instrumentation is not possible. Even in such constrained settings, our \emph{EDG}-based fuzzing can still discover timeouts and crashes in obfuscated binaries, whereas the black-box fuzzer finds none. Our artifact can be found at \url{https://anonymous.4open.science/r/EDG-BBC4}

\bibliographystyle{IEEEtran}
\bibliography{main}

\appendix

\end{document}